\documentclass[prb,twocolumn,showpacs]{revtex4}
\usepackage{amsmath}
\usepackage{amsfonts}
\usepackage{amssymb}
\usepackage{graphicx}

\begin{document}

\title{The Nature of the Superconducting phase Transitions in Strongly
type-II Superconductors in the Pauli Paramagnetic limit}
\author{T. Maniv and V. Zhuravlev}
\affiliation{The Schulich Faculty of Chemistry, Technion - Israel Institute of
Technology, Haifa 32000, Israel. }
\date{\today}

\begin{abstract}
Superconducting phase transitions in strongly type-II superconductors in the
Pauli paramagnetic limit are considered within the framework of the
Gorkov-Ginzburg-Landau approach in the lowest Landau level approximation for
both s and d-wave electron pairing. Simple analytical expressions for the
quadratic and quartic coefficients in the order parameter expansion of the
superconducting free energy are derived without relying on gradient or
wavenumber expansions. The existence of a changeover from continuos to
discontinuos superconducting phase transitions predicted to occur in the
clean limit is shown to depend only on the dimensionality of the underlying
electronic band structure. Such a changeover can take place in the quasi 2D
regime below a critical value of a 3D-2D crossover parameter.
\end{abstract}

\maketitle

\section{Introduction}

In recent years a number of unusual superconducting (SC) states have been
discovered in different new materials \cite{Bennemann04}. Most of these
materials are strongly type-II superconductors, possessing highly
anisotropic or even quasi-two-dimensional (2D) electronic structures. Of
special interest in the present paper are SC materials showing peculiar
clean-limit features at high magnetic fields and low temperatures, notably
the recently discovered family of heavy-fermion compounds CeRIn$_{5}$ (R =
Rh, Ir and Co) \cite{Petrovic01}, and some of the organic charge transfer
salts of the type (BEDT-TTF)$_{2}$X \cite{Ishiguro90},\cite{Wosnitza96}.

The heavy-fermion compound CeCoIn$_{5}$, for example, which is believed to
be an unconventional ( $d$-wave) superconductor \cite{Izawa01} similar to
the high-Tc cuprates, exhibits the highest Tc ($\sim 2.3$ K) among the
Ce-based heavy-Fermion compounds. This material is characterized by
exceptionally strong Pauli paramagnetic pair-breaking \cite{Clogston62},\cite%
{Chandrasekhar62} due to its extremely large electron effective mass and
small Fermi velocity, which could lead to discontinuous (first-order) SC
phase transitions at sufficiently high magnetic fields \cite{Sarma63},\cite%
{Maki64},\cite{Fulde69}.

Recently Bianchi \textit{et.al.}\cite{Bianchi0203} have observed a dramatic
changeover of the second-order SC phase transition to a first-order
transition in specific heat measurements performed on this material as the
magnetic field is increased above some critical values for both parallel and
perpendicular field orientations with respect to the easy conducting planes.
\ Similar effect has been very recently observed by Lortz \textit{et.al. }%
\cite{Lortz07} in the nearly 2D organic superconductor $\kappa $-(BEDT-TTF)$%
_{2}$Cu(NCS)$_{2}$, but only for magnetic field orientation parallel to the
superconducting layers, where the orbital (diamagnetic) pair-breaking is
completely suppressed. Under these conditions the usual (uniform) SC state
is expected to be unstable with respect to formation of a nonuniform SC
state, predicted more than 40 years ago by Fulde and Ferrel \cite{FF64}, and
by Larkin and Ovchinikov \cite{LO64} (FFLO). The corresponding SC order
parameter is spatially-modulated along the field direction with a
characteristic wavenumber, $q$ , whose kinetic energy cost is compensated by
the Pauli pair-breaking energy. The critical temperature, $T_{fflo}$, for
the appearance of the FFLO phase is found to equal $0.56T_{c}$. At the
corresponding tricritical point the normal, the uniform and non-uniform SC
phases are all met.

The possibility of a changeover to first-order transitions can be
effectively investigated within the Ginzburg-Landau (GL) theory of
superconductivity since for the uniform SC phase (i.e. for $q=0$) the
coefficient (usually denoted by $\beta $) of the quartic term in the GL
expansion changes sign at a temperature $T^{\ast }$, which coincides with $%
T_{fflo}$ \cite{Saint-James69}. The identity of $T^{\ast }$ with $T_{fflo}$
is peculiar to the clean limit of a superconductor with no orbital
pair-breaking. In conventional s-wave superconductors electron scattering by
non-magnetic impurities shifts $T_{fflo}$ below the critical temperature $%
T^{\ast }$ \cite{Agterberg01}, allowing discontinuous phase transitions at
temperatures $T_{fflo}<T\leq T^{\ast }$, since (following Anderson's
theorem) $\beta $ is not influenced by nonmagnetic impurities. In
superconductors with unconventional electron pairing, where $\beta $ is
strongly influenced by non-magnetic impurity scattering, the situation is
reversed, i.e. $T^{\ast }<T_{fflo}$.

The interplay between orbital and spin depairing in a pure $s$-wave
isotropic 3D superconductor was first discussed by Gruenberg and Gunther\cite%
{gruenberg66}, who conjectured (i.e. without presenting any result for the
coefficient $\beta $ ) that for $T<0.56T_{c}$ the N-SC transition is of the
second order whereas at lower field there should be a first-order transition
to a uniform SC phase. \ Houzet and Buzdin\cite{houzet01} have essentially
confirmed this picture by exploiting order-parameter and gradient expansions
in the GL theory to find that $T^{\ast }<T_{fflo}$ so that at temperatures $%
T^{\ast }<T<T_{fflo}$, there are second order transitions to either the LO
or FF phase. It should be noted, however, that the orbital effect was
treated there by using gradient expansions, which is a valid approximation
only at very low magnetic fields.

In contrast to all the works outlined above, Adachi and Ikeda have recently
found \cite{adachi03} that, in a clean, $d$-wave, 2D (layered)
superconductor, the orbital effect always shifts $T_{fflo}$ below $T^{\ast }$%
. In this work the authors have used order parameter expansion in the Gorkov
Green's function approach to the GL theory up to six order, avoiding the
restrictions of gradient expansion by exploiting the lowest Landau level
(LLL) approximation for the condensate of Cooper-pairs. Accounting for
impurity scattering destroys the FFLO phase and, in contrast to the pure
paramagnetic situation, somewhat reduces $T^{\ast }$. The effect of SC
thermal fluctuations was found in this work to broaden the discontinuous
mean-field transition at $T^{\ast }$ into a crossover. The reliance on
(FFLO) wavenumber expansion and on extensive numerical computations in this
work has saved formidable analytical efforts, leaving however, interesting
questions unanswered. In particular, the origin of the relative shift of $%
T_{fflo}$ below $T^{\ast }$ by the orbital effect found in this work, in
contrast to all the other works, remains unknown.

In the present paper we develop a formalism based on order parameter
expansion within the Gorkov theory for a strongly type-II superconductor,
with both $s$- and $d_{x^{2}-y^{2}}$-wave electron pairing at high magnetic
fields, which is sufficiently simple to yield useful analytical expressions
for the SC free energy to any desired order in the expansion. The
fundamental interplay between spin induced paramagnetic and orbital
diamagnetic effects at an arbitrary magnetic field is studied, within a
model of anisotropic electron systems covering the entire 3D-2D crossover
range, without relying on gradient or wavenumber expansions. These
advantages enable us to shed new light on the yet undecided debate
concerning the order of the SC phase transitions in the presence of strong
Zeeman spin splitting, and to push our investigation into the unexplored
region of very low temperatures, where quantum magnetic oscillations have
been shown to be observable in the heavy fermion compounds \cite%
{Settai02,Shishido03}.

Specifically, it is found that the relevant parameter controlling the
relative position of $T_{fflo}$ with respect to $T^{\ast }$ is the
dimensionality of the electronic orbital motion in the crystal lattice,
through its influence on the orbital (diamagnetic)\ pair-breaking effect.
For a 3D Fermi surface (isotropic or anisotropic), where the electron motion
along the magnetic field direction reduces the cyclotron kinetic energy, the
shift of $T^{\ast }$ to low temperatures is larger than that of $T_{fflo}$.
In this case the kinetic energy of Cooper-pairs associated with their motion
along the field can compensate the spin-splitting effect, and thus leading
to an increase of $\beta $ and disappearance of the first-order transition.
The corresponding phase diagram is similar to that suggested in Ref. \cite%
{gruenberg66}, where the N-SC transition is of second order, whereas the
transition between non-uniform and uniform (along the field) SC states is of
first order. In the quasi-2D limit (i.e. for quasi-cylindrical Fermi
surfaces) the enhanced orbital pair-breaking shifts $T_{fflo}$ below $%
T^{\ast }$, in agreement with Adachi and Ikeda \cite{adachi03}.

\section{Order parameter expansion in the presence of spin-splitt Landau
levels}

\vspace{1pt}Our starting point is an effective BCS-like Hamiltonian with a $%
d_{x^{2}-y^{2}}$-wave pairing interaction similar to that exploited,e.g. by
Agterberg and Yang \cite{Agterberg01}. The conventional $s$-wave situation
can be similarly worked out and so will not be presented in detail here. The
thermodynamical potential (per unit volume) for the corresponding $d$-wave
superconductor, as expanded in the order parameter with nonlocal normal
electron kernels, may be written as: 
\begin{equation}
\Omega =\frac{\Delta _{0}^{2}}{V}+\sum\limits_{m=1}\frac{\left( -1\right)
^{m}}{m}\widetilde{\Omega }_{2m}\left\{ \Delta \left( \mathbf{R},\mathbf{r}%
\right) \right\}  \label{OPExpan}
\end{equation}%
where $\widetilde{\Omega }_{2m}\left\{ \Delta \left( \mathbf{R},\mathbf{r}%
\right) \right\} $ is a functional of the SC order parameter, $\Delta \left( 
\mathbf{R},\mathbf{r}\right) $, having a power-low dependence $\sim |\Delta
_{0}|^{2m}$ on the global amplitude, $\Delta _{0}$ , of the order parameter,
and $V$ is a BCS coupling constant (given in units of energy$\times $
volume). The corresponding $d$-wave order parameter depends on both the
center of mass ($\mathbf{R}$) and relative ($\mathbf{r}$) coordinates of a
condensate of electron pairs: $\Delta (\mathbf{R},\mathbf{r})=\Delta (%
\mathbf{R})\varphi (\mathbf{r})$. It should be determined self-consistently
from the corresponding pair-correlation functions. Only stationary solutions
are considered, neglecting quantum and thermal fluctuations. In addition the
order parameter in the mean field approximation is selected as a hexagonal
vortex lattice. Actually this assumption is not very important since the
second order term in the order parameter expansion does not depend on the
vortex lattice structure whereas the lattice structure dependence of the
quartic term is very weak (for reviews see \cite{Rasolt-Tesanovic92},\cite%
{MRVW92}).

For the underlying system of normal electrons we assume a simple model of
quadratic energy dispersion $\varepsilon \left( k_{x},k_{y},k_{z}\right)
=\hbar ^{2}\left( k_{x}^{2}+k_{y}^{2}\right) /2m^{\ast }+\hbar
^{2}k_{z}^{2}/2m_{z}^{\ast }$ and anisotropic effective mass tensor: $%
m^{\ast }\leq m_{z}^{\ast }$. A quasi-2D situation is characterized by a
sufficiently large anisotropy parameter $\chi _{a}=\sqrt{m_{z}^{\ast
}/m^{\ast }}$, corresponding to an elongated Fermi surface with a Fermi
momentum $k_{F}$ and Fermi energy $\varepsilon _{F}\equiv \hbar
^{2}k_{F}^{2}/2m^{\ast }$, which is truncated by the Brillouin zone (BZ)
face at $k_{z,\max }=\pi /d$ , where $d$ is the lattice constant
perpendicular to the easy planes. A parameter determining the dimensionality
of the Fermi surface may be defined by: $v_{0}=\sqrt{\frac{\varepsilon
_{z,\max }}{\varepsilon _{F}}}$ , where $\varepsilon _{z,\max }\equiv \hbar
^{2}k_{z,\max }^{2}/2m_{z}^{\ast }$ is the maximal value of the electron
energy along the field. Thus, in the 2D limit, $\ k_{z,\max }\ll k_{F}$ , we
have $v_{0}\rightarrow 0$ , while the system may be regarded 3D (isotropic
or anisotropic ) if $\ k_{z,\max }\simeq k_{F}$ for which the Fermi surface
is contained entirely within the first BZ, namely for $v_{0}=1$. \ 

At any order of the expansion, Eq.(\ref{OPExpan}), the nonlocal electronic
kernel of the corresponding functional (see e.g. Eqs. (\ref{Omega_gen}),(\ref%
{Kernel_2}), and Eqs. (\ref{Omega4_def}),(\ref{K4_def})) consists of a
product of $m=1,2,...$ pairs of normal electron Green's functions in a
constant magnetic field,$\mathbf{H}=H\widehat{z}$ (i.e. perpendicular to the
easy conducting layers), which are written in the form: \ $G_{\uparrow
\downarrow }\left( \mathbf{R}_{1},\mathbf{R}_{2},\omega _{\nu }\right)
=G_{0\uparrow \downarrow }\left( \mathbf{R}_{2}-\mathbf{R}_{1},\omega _{\nu
}\right) g\left( \mathbf{R}_{1},\mathbf{R}_{2}\right) $, where the gauge
factor is given by$\ g\left( \mathbf{R}_{1},\mathbf{R}_{2}\right) =e^{-\frac{%
i}{2a_{H}}\left[ \mathbf{R}_{1}\times \mathbf{R}_{2}\right] \cdot \widehat{z}%
}$, $a_{H}=\sqrt{c\hbar /eH}$ , and the gauge invariant part can be
calculated by the well known expression\cite{Bychkov62} 
\begin{widetext}
\begin{equation}
G_{0\uparrow \downarrow }\left( \mathbf{R}_{2}-\mathbf{R}_{1},\omega _{\nu
}\right) =\frac{1}{2\pi a_{H}^{2}}\int \frac{dk_{z}}{2\pi }%
e^{ik_{z}(Z_{2}-Z_{1})}\sum_{n}\dfrac{e^{-\rho ^{2}/4}L_{n}(\rho ^{2}/2)}{%
\mu -\varepsilon _{nk_{z}\uparrow \downarrow }+i\hbar \omega _{\nu }+i%
\mathrm{sign}\left( \omega _{\nu }\right) \hbar \Gamma }  \label{GreenFunc}
\end{equation}

Here $\omega _{\nu }=\pi k_{B}T\left( 2\nu +1\right) /\hbar $ with $\nu
=0,\pm 1,...$ is Matzubara frequency, $\varepsilon _{nk_{z}\uparrow }=\hbar
\omega _{c}\left( n+1/2+x^{2}-g/2\right) $ , \ $\varepsilon
_{nk_{z}\downarrow }=\omega _{c}\left( n+1/2+x^{2}+g/2\right) $, the
spin-split normal electron energy levels, $\omega _{c}=eH/m^{\ast }$-the
in-plane electronic cyclotron frequency, $x^{2}=\xi ^{2}k_{z}^{2}\equiv 
\frac{k_{z}^{2}}{2m_{z}^{\ast }\omega _{c}}$is a dimensionless longitudinal
(parallel to the magnetic field) kinetic energy, $\omega _{c}g\equiv
eH/m_{0} $, is the Zeeman spin splitting energy, $\Gamma $- the impurity
scattering relaxation rate, and $\mu =\hbar \omega _{c}\left(
n_{F}+1/2\right) \approx \varepsilon _{F}=\hbar ^{2}k_{F}^{2}/2m^{\ast }$ is
the chemical potential. The spatial variables are dimensionless in-plane
(perpendicular to the magnetic field) coordinates, $\mathbf{\rho }=\frac{%
\mathbf{R}_{2\bot }-\mathbf{R}_{1\bot }}{a_{H}}$, and longitudinal
coordinates:$\ Z_{1}=\mathbf{R}_{1}\cdot \widehat{z}$ , and $Z_{2}=\mathbf{R}%
_{2}\cdot \widehat{z}$. \ \ 

\subsection{The quadratic term}

In the expansion, Eq.(\ref{OPExpan}), the second order term, which describes
the SC condensation energy of spin-singlet electron-pairs, propagating from
initial ($i=1$) to final ($i=2$) coordinates $\mathbf{R}_{i}\pm \frac{%
\mathbf{r}_{i}}{2}$, is given by: 
\begin{equation}
\Omega _{2}=\frac{\Delta _{0}^{2}}{V}-\frac{1}{\mathcal{V}_{0}}\int d^{3}%
\mathbf{R}_{1}d^{3}\mathbf{R}_{2}\tilde{\Gamma}_{2}\left( \mathbf{R}_{1},%
\mathbf{R}_{2}\right) \tilde{K}_{2}\left( \mathbf{R}_{1},\mathbf{R}%
_{2}\right) \equiv \frac{\Delta _{0}^{2}}{V}-A_{0}\Delta _{0}^{2}
\label{Omega_gen}
\end{equation}%
where $\mathcal{V}_{0}=SL_{z}$ is the volume of the system. The vertex part, 
$\tilde{\Gamma}_{2},$ is a product of two order parameters multiplied by the
gauge factors, $g\left( \mathbf{R}_{2},\mathbf{R}_{1}\right) $, which are
functions of the center of mass coordinates only, due to cacellation by the
corresponding phase factors of the order parameters, namely:%
\begin{equation}
\tilde{\Gamma}_{2}\left( \mathbf{R}_{1},\mathbf{R}_{2}\right) =g^{\ast
}\left( \mathbf{R}_{1},\mathbf{R}_{2}\right) g\left( \mathbf{R}_{2},\mathbf{R%
}_{1}\right) \Delta \left( \mathbf{R}_{1}\right) \Delta ^{\star }\left( 
\mathbf{R}_{2}\right)  \label{Vertex_def}
\end{equation}%
The kernel $\tilde{K}_{2}$ is a product of two translational invariant
Green's functions, convoluted with the corresponding factors of the order
parameters, which depend only on the relative pair coordinates, namely: 
\begin{eqnarray}
&&\tilde{K}_{2}\left( \mathbf{R}_{1},\mathbf{R}_{2}\right) =k_{B}T\sum_{\nu
}\int d^{3}\mathbf{r}_{1}d^{3}\mathbf{r}_{2}\varphi \left( \mathbf{r}%
_{1}\right) \varphi ^{\ast }\left( \mathbf{r}_{2}\right)  \label{Kernel_2} \\
&&\times G_{0\uparrow }^{\ast }\left( \mathbf{R}_{2}-\mathbf{R}_{1}+\frac{%
\mathbf{r}_{2}-\mathbf{r}_{1}}{2},\omega _{\nu }\right) G_{0\downarrow
}\left( \mathbf{R}_{1}-\mathbf{R}_{2}+\frac{\mathbf{r}_{2}-\mathbf{r}_{1}}{2}%
,\omega _{\nu }\right)  \notag
\end{eqnarray}

The factor of the order parameter which depends on the pair center of mass
coordinates is written as\cite{RMP01}:%
\begin{equation}
\Delta \left( \mathbf{R}\right) =c(Z)\sum\limits_{n}e^{i\pi n^{2}/2}\phi
_{n}\left( \mathbf{R}_{\bot }\right) ;\ \ \ \ \ \ \ \ \phi _{n}\left( 
\mathbf{R}_{\bot }\right) =e^{i\frac{2\pi n}{a_{x}}X-\left( Y-\frac{\pi n}{%
a_{x}}\right) ^{2}};\ \ \ \ \ \ a_{x}=  \label{OrderParam}
\end{equation}%
where $c(Z)=c_{0}e^{iqZ}$ is the Fulde-Ferrell modulation factor. Exploiting
the fact that the kernel $\tilde{K}_{2}\left( \mathbf{R}_{1},\mathbf{R}%
_{2}\right) $ depends only on the difference $\mathbf{R}_{1}-\mathbf{R}_{2}$%
, one may carry out the integration in Eq.(\ref{Omega_gen}) first over the
in-plane mean coordinates $\mathbf{R}_{\perp }\mathbf{=}\left( \mathbf{R}%
_{\bot ,1}+\mathbf{R}_{\bot ,2}\right) /2$ to get the following average
vertex part \cite{Stephen92} 
\begin{equation}
\left\langle \tilde{\Gamma}_{2}\right\rangle =\frac{1}{V}\int \tilde{\Gamma}%
_{2}\left( \mathbf{R}_{1},\mathbf{R}_{2}\right) d^{2}R_{\bot }=|c_{0}|^{2}%
\frac{a_{x}}{\sqrt{2\pi }}e^{-\rho ^{2}/2-iq(Z_{2}-Z_{1})}=\Delta
_{0}^{2}e^{-\rho ^{2}/2-iq(Z_{2}-Z_{1})}  \label{Vertex_22}
\end{equation}%
where $\Delta _{0}^{2}=\frac{1}{\mathcal{V}_{0}}\int d^{3}R|\Delta \left( 
\mathbf{R}\right) |^{2}$ ($\mathcal{V}_{0}=SL_{z}$), and then integrate over
the rest of the coordinates $\mathbf{\rho =}\left( \mathbf{\mathbf{R}_{\bot
,2}}-\mathbf{R}_{\bot ,1}\right) /a_{H}$\ and $\rho _{z}=\left(
Z_{2}-Z_{1}\right) /a_{H}$.

Since, among other things, we are interested in the effect of quantum
magnetic oscillations, we apply a technique of exact summation over LLs
suggested in Ref.\cite{hait}. It is similar to the Poisson summation
formula, which transforms the summation over LLs into summation over
harmonics of the inverse magnetic field, and allows to deal seperately with
the uniform (quasi-classical) contribution and the various quantum
corrections. This technique can be briefly described as follows. Let us
consider the integral representation of the Green's functions, $\left[
n_{F}-n-x^{2}\pm i\omega \right] ^{-1}=\int_{0}^{\infty }d\tau e^{\pm i\tau %
\left[ n_{F}-n-x^{2}\pm i\omega \right] }$ and perform the summation over
LLs using the well known identity, $\sum_{n=0}^{\infty }z^{n}L_{n}\left(
t\right) =\left( 1-z\right) ^{-1}\exp \left( \frac{tz}{z-1}\right) $ with $%
z=e^{\pm i\tau }$ and $t=\rho ^{2}/2$. Taking advantage of these relations
the gauge invariant part of the Green's function for $\omega _{\nu }\geq 0$
can be transformed to:%
\begin{equation}
G_{0\uparrow \downarrow }\left( \mathbf{R}_{2}-\mathbf{R}_{1};\omega _{\nu
}\right) =\frac{1}{2\pi a_{H}^{2}\hbar \omega _{c}}\int \frac{dk_{z}}{2\pi }%
e^{ik_{z}(Z_{2}-Z_{1})}\int_{0}^{\infty }d\tau \frac{e^{i\tau \left[
n_{F}-x^{2}+g+i\widetilde{\omega }_{\nu }+i\widetilde{\Gamma }\right] }}{%
\left( 1-e^{-i\tau }\right) }\exp \left( \frac{\rho ^{2}}{4}\frac{1+e^{i\tau
}}{1-e^{i\tau }}\right)  \label{GreenFunc_tau}
\end{equation}%
where $\widetilde{\omega }_{\nu }\equiv \omega _{\nu }/\omega _{c}$, $%
\widetilde{\Gamma }\equiv \Gamma /\omega _{c}$. For $\omega _{\nu }<0$ one
should replace $\tau $ with $-\tau $ (or $\omega _{\nu }$ with $-|\omega
_{\nu }|$).

The scattering of electrons by non-magnetic impurities is taken into account
here as a self-energy correction to the single electron Green's functions
using the standard relaxation time approximation. Vertex corrections to the
quadratic kernel $\tilde{K}_{2}\left( \mathbf{R}_{1},\mathbf{R}_{2}\right) $
(as well as to higher order ones), which are known to exactly cancel the
self-energy insertions in the very weak magnetic field regime of
convensional s-wave superconductors (see e.g. \cite{Werthamer69}), are not
so crucial in the strong magnetic field regime of both the s and d-wave
situations investigated here, and will be therefore neglected in our
calculations, as done, e.g. in Refs.\cite{adachi03,Mineev}. In any event,
for the high magnetic field and relatively clean superconductors considered
here, the length scale, $a_{H}$, corresponding to the diamagnetic
pair-breaking is much smaller than the electron mean free path $v_{F}/\Gamma 
$, and the effect of impurity scattering is marginal.

Utilizing this approximation we rewrite the kernel in the following form: 
\begin{eqnarray}
\tilde{K}_{2}\left( \mathbf{\rho },\rho _{z}\right) &=&\frac{k_{B}T}{\left(
2\pi a_{H}^{2}\hbar \omega _{c}\right) ^{2}}\sum_{\nu }\int dz_{1}dz_{2}\int 
\frac{dk_{z,1}}{2\pi }e^{ik_{z,1}(\rho _{z}+\sigma _{z}/2)}\int \frac{%
dk_{z,2}}{2\pi }e^{ik_{z,2}(\rho _{z}-\sigma _{z}/2)}  \label{Kernel2} \\
&&\int_{0}^{\infty }d\tau _{1}e^{i\tau _{1}\left[ n_{F}-\xi
^{2}k_{z,1}^{2}+g+i\widetilde{\omega }_{\nu }+i\widetilde{\Gamma }\right]
}\int_{0}^{\infty }d\tau _{2}e^{-i\tau _{2}\left[ n_{F}-\xi
^{2}k_{z,2}^{2}-g-i\widetilde{\omega }_{\nu }-i\widetilde{\Gamma }\right]
}J\left( \tau _{1},\tau _{2,}\mathbf{\rho }\right)  \notag
\end{eqnarray}%
where $\ $ $\mathbf{\sigma =}\left( \mathbf{r}_{\perp ,2}-\mathbf{r}_{\perp
,1}\right) /a_{H}$ , $\sigma _{z}=\left( z_{2}-z_{1}\right) /a_{H}$,\ and%
\begin{eqnarray}
J\left( \tau _{1},\tau _{2,}\mathbf{\rho }\right) &=&\int d^{2}\mathbf{r}%
_{\perp ,1}d^{2}\mathbf{r}_{\perp ,2}f^{\ast }\left( \mathbf{r}_{2}\right)
\left( 1-e^{-i\tau _{1}}\right) ^{-1}\exp \left( \frac{\left( \mathbf{\rho }+%
\mathbf{\sigma }/2\right) ^{2}}{4}\frac{1+e^{i\tau _{1}}}{1-e^{i\tau _{1}}}%
\right)  \notag \\
&&\times \left( 1-e^{i\tau _{2}}\right) ^{-1}\exp \left( \frac{\left( 
\mathbf{\rho }-\mathbf{\sigma }/2\right) ^{2}}{4}\frac{1+e^{-i\tau _{2}}}{%
1-e^{-i\tau _{2}}}\right)  \label{Jdef}
\end{eqnarray}

In Eq.(\ref{Kernel2}) we use the representation $\varphi \left( \mathbf{r}%
\right) =\delta \left( z\right) f\left( \mathbf{r}_{\perp }\right) $ , where 
$f\left( \mathbf{r}_{\perp }\right) =\int \frac{d^{2}k}{\left( 2\pi \right)
^{2}}f_{\mathbf{k}}e^{i\left( \mathbf{k\cdot r}_{\perp }\right) }$ describes
the two types of the electron pairing, the symmetric $s$-wave pairing with%
\begin{equation}
f_{sk}=\frac{1}{2}\left( \cos \left( k_{x}d\right) +\cos \left(
k_{y}d\right) \right)  \label{s_pair}
\end{equation}%
and $d_{x^{2}-y^{2}}$-wave pairing,%
\begin{equation}
f_{dk}=\frac{1}{2}\left( \cos \left( k_{x}d\right) -\cos \left(
k_{y}d\right) \right)  \label{d_pair}
\end{equation}%
producing nodes in the order parameter along the $k_{x}=\pm k_{y}$
directions. The $\delta $-dependence on $z$ enables us to readily perfom the
first two integrations in Eq. (\ref{Kernel2}).

Using the resulting expression for $\tilde{K}_{2}\left( \mathbf{\rho },\rho
_{z}\right) $, and the vertex function $\left\langle \tilde{\Gamma}%
_{2}\right\rangle $ one can calculate the nontrivial coefficient $A_{0}$ in
Eq.(\ref{Omega_gen}) by performing the integrals over $\mathbf{\rho }$,and $%
\rho _{z}$ 
\begin{equation}
A_{0}=\int d^{2}\mathbf{\rho }d\rho _{z}\tilde{K}_{2}\left( \mathbf{\rho }%
,\rho _{z}\right) e^{-\rho ^{2}/2-iq\rho _{z}}  \label{A0_rho}
\end{equation}%
with the other integrals incorporated in the kernel as appearing in Eq.(\ref%
{Kernel2}). It is convenient to perform the integration over $\mathbf{\rho }$
first since both the function $J\left( \tau _{1},\tau _{2,}\mathbf{\rho }%
\right) $and the vertex part have a gaussian dependence on $\mathbf{\rho }$
which can be readily carried out with the result:%
\begin{equation}
J\left( \tau _{1},\tau _{2}\right) =\int d^{2}\mathbf{\rho }e^{-\rho
^{2}/2}J\left( \tau _{1},\tau _{2,}\mathbf{\rho }\right) =\frac{2\pi
J_{p}\left( \tau _{1},\tau _{2}\right) }{2-e^{-i\tau _{1}}-e^{i\tau _{2}}}
\label{Jp1}
\end{equation}%
where%
\begin{equation}
J_{p}\left( \tau _{1},\tau _{2}\right) =\int d^{2}\mathbf{r}_{\perp ,1}d^{2}%
\mathbf{r}_{\perp ,2}f\left( \mathbf{r}_{1}\right) f^{\ast }\left( \mathbf{r}%
_{2}\right) e^{-\frac{\gamma _{\tau }}{8}\mathbf{\sigma }^{2}}  \label{Jp2}
\end{equation}%
and $\gamma _{\tau }=\frac{2+e^{-i\tau _{1}}+e^{i\tau _{2}}}{2-e^{-i\tau
_{1}}-e^{i\tau _{2}}}$. \ It should be noted here that the type of pairing
influences the SC condensation energy through the functional dependence of $%
J_{p}$ on the pairing function $f_{\mathbf{k}}$.

Performing the straightforward calculation of $J_{p}\left( \tau _{1},\tau
_{2}\right) $ for both functions one obtains,%
\begin{equation}
J_{sp}\left( \tau _{1},\tau _{2}\right) =\frac{1}{4}\left( 1+e^{-\frac{1}{4}%
\gamma _{\tau }d^{2}}\right) ^{2}\simeq 1  \label{Jsp}
\end{equation}%
\begin{equation}
J_{dp}\left( \tau _{1},\tau _{2}\right) =\frac{1}{4}\left( 1-e^{-\frac{1}{4}%
\gamma _{\tau }d^{2}}\right) ^{2}\simeq \frac{1}{4}\left( \frac{\gamma
_{\tau }d^{2}}{4}\right) ^{2}  \label{Jdp}
\end{equation}%
where the last approximate step is obtained in the limit $\frac{\gamma
_{\tau }}{4}d^{2}\ll 1.$ \ This can be justified by noting that the scale of
the function $\gamma _{\tau }$ is of the order unity whereas $d$ (in units
of the magnetic length) is much smaller than one. In the opposite limit: $%
J_{sp}\left( \tau _{1},\tau _{2}\right) =J_{dp}\left( \tau _{1},\tau
_{2}\right) =\frac{1}{4}$\textbf{.}

Thus, noting that the integration of\textbf{\ }$A_{0}$\textbf{\ }over the
center of mass coordinates yields just the total volume of the system, and
performing the integration\ over the relative coordinate $\rho _{z}$,%
\begin{eqnarray}
&&\int d\rho _{z}\int \frac{dk_{z,1}}{2\pi }e^{ik_{1,z}\rho _{z}}\int \frac{%
dk_{z,2}}{2\pi }e^{ik_{z,2}\rho _{z}}\ e^{i\tau _{1}\left[ n_{F}-\xi
^{2}k_{z,1}^{2}+g+i\widetilde{\omega }_{\nu }+i\widetilde{\Gamma }\right]
}e^{-i\tau _{2}\left[ n_{F}-\xi ^{2}k_{z,2}^{2}-g-i\widetilde{\omega }_{\nu
}-i\widetilde{\Gamma }\right] }e^{-iq\rho _{z}}  \notag \\
&=&\int \frac{dk_{z}}{2\pi }\ e^{i\tau _{1}\left[ n_{F}-\xi ^{2}\left(
k_{z}+q/2\right) ^{2}+g+i\widetilde{\omega }_{\nu }+i\widetilde{\Gamma }%
\right] }e^{-i\tau _{2}\left[ n_{F}-\xi ^{2}\left( k_{z}-q/2\right) ^{2}-g-i%
\widetilde{\omega }_{\nu }-i\widetilde{\Gamma }\right] },  \label{Kz_integ}
\end{eqnarray}%
one obtaines for a $d$-wave superconductor:%
\begin{eqnarray}
A_{0}^{\left( d\right) } &=&\frac{6k_{B}T}{\left( \hbar ^{2}k_{z,\max
}^{2}/2m^{\ast }\right) ^{2}a_{H}^{2}}\sum_{\nu }\int \frac{dk_{z}}{2\pi }%
\int_{-\infty }^{\infty }d\tau _{1}d\tau _{2}\   \label{A0_d1} \\
&&\times \frac{e^{i\tau _{1}\left[ n_{F}-\xi ^{2}\left( k_{z}+q/2\right)
^{2}+g+i\widetilde{\omega }_{\nu }+i\widetilde{\Gamma }\right] }e^{-i\tau
_{2}\left[ n_{F}-\xi ^{2}\left( k_{z}-q/2\right) ^{2}-g-i\widetilde{\omega }%
_{\nu }-i\widetilde{\Gamma }\right] }}{\left( 2-e^{-i\tau _{1}}-e^{i\tau
_{2}}\right) ^{3}}.  \notag
\end{eqnarray}

A similar expression can be derived for an $s$-wave superconductor. Below we
will present only the final result for this case (\emph{\ }see Eqs.\textbf{\ 
}\ref{A0_fin}\textbf{,}\ref{Theta_s}\textbf{)}.

Eq. (\ref{A0_d1}) is an \textit{exact} representation for the coefficient of
the quadratic term in the order parameter expansion, Eq.\ref{Omega_gen},
which includes low temperature quantum corrections and quantum magnetic
oscillations. It can be written as a sum of contributions from poles at the
2D lattice: $\tau _{1}=2\pi n_{1}$ and $\tau _{2}=2\pi n_{2}$ , with $%
n_{1,2}=0,1,...$. The dominant (zero harmonic) quasiclassical contribution
arises from the pole at $n_{1}=n_{2}=0$ whereas the quantum corrections are
associated with the poles at $n_{1}=n_{2}\not=0$. It is easy to see that the
oscillating terms correspond to the off-diagonal poles, $n_{1}\not=n_{2}$,
in the $\left( \tau _{1},\tau _{2}\right) $-plane.

In the present paper we are interested mainly in the quasiclassical
contribution for which a further simplification can be achieved. Changing to
new variables: $\tau _{2}=\rho _{0}+\frac{\tau }{2};$ \ \ \ $\tau _{1}=\rho
_{0}-\frac{\tau }{2}$, and exploiting the expansion $2-e^{-i\tau
_{1}}-e^{i\tau _{2}}\simeq \rho _{0}^{2}-i\tau $ near the "quasiclassical"
pole $\tau _{1}=\tau _{2}=0$ , one carries out the integral over $\tau $ in
Eq. (\ref{A0_d1}) to have:%
\begin{eqnarray}
A_{0}^{\left( d\right) } &=&\frac{6k_{B}T}{\left( \hbar ^{2}k_{z,\max
}^{2}/2m^{\ast }\right) ^{2}a_{H}^{2}}\sum_{\nu }\int dk_{z}\int_{0}^{\infty
}d\rho _{0}e^{2i\rho _{0}\left[ \xi ^{2}qk_{z}+g+i\widetilde{\omega }_{\nu
}+i\widetilde{\Gamma }\right] }  \notag \\
&&\times \left[ n_{F}-\xi ^{2}\left( k_{z}^{2}+\left( q/2\right) ^{2}\right) %
\right] ^{2}e^{-\rho _{0}^{2}\left[ n_{F}-\xi ^{2}\left( k_{z}^{2}+\left(
q/2\right) ^{2}\right) \right] }  \label{A0_d2}
\end{eqnarray}%
Note that the lowest order expansion of the denominator in Eq.(\ref{A0_d1})
about $\tau _{1}=\tau _{2}=0$ is kept under the entire range of integration
since the important integration inerval is of the order $\tau \sim \rho
_{0}^{2}\ll \rho _{0}\sim \frac{1}{\sqrt{n_{F}}}\ll 1$. Note also that
throughout this paper we assume that $n_{F}\gg 1$.

It is convenient to rescale variables as%
\begin{equation}
u\equiv \sqrt{n_{F}}\rho _{0};\ \ \ x_{0}\equiv \xi q;\ \ \ v\equiv \frac{%
\xi k_{z}}{\sqrt{n_{F}}}  \label{scale_var}
\end{equation}%
and neglect the energy of an electron pair along the $z$-axis, $\left( \xi
q\right) ^{2}$, with respect to Fermi energy, $n_{F}$. Perfoming the
explicit summation over Matsubara frequencies one obtains, in terms of the
new variables, the following result:%
\begin{equation}
A_{0}^{\left( d\right) }=N\left( 0\right) \lambda _{d}\frac{2\pi k_{B}T}{%
\sqrt{\mu \hbar \omega _{c}}}\int_{0}^{\infty }du\frac{1-e^{-\frac{2\omega
_{D}}{\sqrt{\mu \hbar \omega _{c}}}u}}{\sinh \left( \frac{2\pi k_{B}T}{\sqrt{%
\mu \hbar \omega _{c}}}u\right) }e^{-\frac{2\widetilde{\Gamma }}{\sqrt{n_{F}}%
}u}\cos \left( \frac{2g}{\sqrt{n_{F}}}u\right) \Theta _{2}^{\left( d\right)
}\left( u,x_{0}\right)  \label{A0_fin}
\end{equation}%
where $\lambda _{d}=3\left( \frac{k_{F}d}{\pi }\right) ^{4}$, $N(0)$ is the
electron density of states per spin at the Fermi energy\texttt{( }$N(0)=%
\frac{\sqrt{m^{\ast }m_{z}^{\ast }}k_{F}}{2\pi \hbar ^{3}}\ $), and: 
\begin{equation}
\Theta _{2}^{\left( d\right) }\left( u,x_{0}\right) =\int_{0}^{1}d\nu \left(
1-v^{2}\right) ^{2}\cos \left( 2x_{0}uv\right) e^{-u^{2}\left(
1-v^{2}\right) }.  \label{Theta_d}
\end{equation}

A similar result is obtained for an $s$-wave superconductor. In this case $%
\lambda _{d}\rightarrow \lambda _{s}=1$ and 
\begin{equation}
\Theta _{2}^{\left( s\right) }\left( u,x_{0}\right) =\int_{0}^{1}d\nu \cos
\left( 2x_{0}uv\right) e^{-u^{2}\left( 1-v^{2}\right) }.  \label{Theta_s}
\end{equation}%
For $g=x_{0}=\widetilde{\Gamma }=0$ Eqs. (\ref{A0_fin})(\ref{Theta_s})
reduces to the quadratic term derived by Helfand-Werthammer \cite{WHH}.

\subsection{The quartic term}

The quartic term in the perturbation expansion, Eq.(\ref{OPExpan}), which
corresponds to a closed loop diagram with four vertices, is given by:%
\begin{equation}
\Omega _{4}^{(s,d)}=\frac{1}{\mathcal{V}_{0}}\int d^{3}\mathbf{R}_{1}d^{3}%
\mathbf{R}_{2}d^{3}\mathbf{R}_{3}d^{3}\mathbf{R}_{4}\widetilde{\Gamma }%
_{4}\left( \mathbf{R}_{1},\mathbf{R}_{2},\mathbf{R}_{3},\mathbf{R}%
_{4}\right) \tilde{K}_{4}\left( \mathbf{R}_{1},\mathbf{R}_{2},\mathbf{R}_{3},%
\mathbf{R}_{4}\right) ,  \label{Omega4_def}
\end{equation}%
where the kernel, containing the gauge invariant factors of the four
electron Green's functions, is: 
\begin{equation*}
\tilde{K}_{4}\left( \mathbf{R}_{1},\mathbf{R}_{2},\mathbf{R}_{3},\mathbf{R}%
_{4}\right) =k_{B}T\sum_{\nu }\int d^{3}\mathbf{r}_{1}d^{3}\mathbf{r}%
_{2}d^{3}\mathbf{r}_{3}d^{3}\mathbf{r}_{4}\varphi \left( \mathbf{r}%
_{1}\right) \varphi ^{\ast }\left( \mathbf{r}_{2}\right) \varphi \left( 
\mathbf{r}_{3}\right) \varphi ^{\ast }\left( \mathbf{r}_{4}\right)
\end{equation*}%
\begin{eqnarray}
&&\times G_{0\uparrow }^{\ast }\left( \mathbf{R}_{2}-\mathbf{R}_{1}+\frac{%
\mathbf{r}_{2}-\mathbf{r}_{1}}{2},\omega _{\nu }\right) G_{0\downarrow
}\left( \mathbf{R}_{3}-\mathbf{R}_{2}-\frac{\mathbf{r}_{3}-\mathbf{r}_{2}}{2}%
,\omega _{\nu }\right)  \notag \\
&&\times G_{0\uparrow }^{\ast }\left( \mathbf{R}_{4}-\mathbf{R}_{3}+\frac{%
\mathbf{r}_{4}-\mathbf{r}_{3}}{2},\omega _{\nu }\right) G_{0\downarrow
}\left( \mathbf{R}_{1}-\mathbf{R}_{4}-\frac{\mathbf{r}_{1}-\mathbf{r}_{4}}{2}%
,\omega _{\nu }\right)  \label{K4_def}
\end{eqnarray}%
and the vertex part: 
\begin{equation}
\widetilde{\Gamma }_{4}\left( \mathbf{R}_{1},\mathbf{R}_{2},\mathbf{R}_{3},%
\mathbf{R}_{4}\right) =g^{\ast }\left( \mathbf{R}_{1},\mathbf{R}_{2}\right)
g\left( \mathbf{R}_{2},\mathbf{R}_{3}\right) g^{\ast }\left( \mathbf{R}_{3},%
\mathbf{R}_{4}\right) g\left( \mathbf{R}_{4},\mathbf{R}_{1}\right) \Delta
\left( \mathbf{R}_{1}\right) \Delta ^{\ast }\left( \mathbf{R}_{2}\right)
\Delta \left( \mathbf{R}_{3}\right) \Delta ^{\ast }\left( \mathbf{R}%
_{4}\right)  \label{Gamma4_def}
\end{equation}%
which consists of the gauge factors $g(\mathbf{R}_{i},\mathbf{R}_{j})$ and
the order parameter values at the four center of mass positions for two
electron pairs. \ 

Since the dependence of the order parameter on the relative pair coordinates
is separable from that of the center of mass coordinates, the latter
dependence is selected to have the usual Abrikosov lattice structure,%
\begin{equation}
\Delta \left( \mathbf{R}\right) =c_{0}\ e^{iqZ}e^{-\frac{1}{2}\left(
|u|^{2}-u^{2}\right) }\sum_{n=0,\pm 1,\pm 2,...}e^{iq_{n}u-q_{n}^{2}/4},
\label{Delta_comp}
\end{equation}%
with $q_{n}=2\pi n/a_{x}$ \ and $u=X+iY$, . To simplfy the calculation of
the vertex part we exploit several assumptions. Substituting Eq.(\ref%
{Delta_comp}) to Eq.(\ref{Gamma4_def}), one may keep only diagonal terms
with $q_{n1}=q_{n2}=q_{n3}=q_{n4}=p$ , since all off-diagonal terms are
small by the gaussian factor $\sim \exp \left[ -\left( q_{n4}-q_{n1}\right)
^{2}-\left( q_{n4}-q_{n1}\right) ^{2}\right] $. Furthermore, we may replace
summation over $p$ with an appropriate integration. Both of these
assumptions are equivalent to neglecting particular vortex lattice
structures, corresponding to replacement of the Abrikosov structure
parameter, $\beta _{A}$, with $\frac{\sqrt{\pi }}{a_{x}}$ \cite{RMP01},
which yields only a small error.

With the above assumptions the vertex part reduces to: 
\begin{equation}
\widetilde{\Gamma }_{4}\left( \mathbf{R}_{1},\mathbf{R}_{2},\mathbf{R}_{3},%
\mathbf{R}_{4}\right) =\frac{a_{x}\sqrt{\pi }}{2\pi }|c_{0}|^{4}e^{iq\left(
Z_{1}-Z_{2}+Z_{3}-Z_{4}\right) }e^{-\frac{1}{4}\sum_{l=1}^{3}|\rho
_{l}|^{2}}e^{\frac{1}{4}\left[ \left( u_{1}-u_{3}\right) ^{2}+\left(
u_{2}^{\ast }-u_{4}^{\ast }\right) ^{2}\right] }  \label{Gamma4_comp}
\end{equation}%
where $\mathbf{\rho }_{l}=u_{l+1}-u_{l}$. Since the dominant contribution to
the quartic term arises from small propagation distances, $|u_{l}|\ \leq 1$%
\cite{MRVW92,ZM97}, one may expand the last exponential on the RHS of Eq.(%
\ref{Gamma4_comp}), up to leading order, under the integrals over angular
variables in Eq.(\ref{Omega4_def}). Additional angular dependence is due to
the kernel, $K_{4}$, through its dependence on the absolute values of linear
combinations of "external", $\mathbf{R}_{l+1}-\mathbf{R}_{l}$, and
"internal", $\mathbf{r}_{l+1}-\mathbf{r}_{l}$ ($l=1,..,4$), coordinates (see
Eq. \ref{K4_def}). Since the characteristic size of $\left\vert \mathbf{r}%
_{l+1}-\mathbf{r}_{l}\right\vert \sim d$ is much smaller than the scale of $%
\left\vert \mathbf{R}_{l+1}-\mathbf{R}_{l}\right\vert \sim a_{H}$ , the
dependence of the kernel on $\mathbf{r}_{l+1}-\mathbf{r}_{l}$ (and
consequently its dependence on the angular variables) may be neglected at
large $\left\vert \mathbf{R}_{l+1}-\mathbf{R}_{l}\right\vert $. \ Therefore,
the integration over angular variables in this region involves only the last
exponential in Eq.(\ref{Gamma4_comp}), resulting in: 
\begin{eqnarray*}
\left\langle e^{\frac{1}{4}\left[ \left( u_{1}-u_{3}\right) ^{2}+\left(
u_{2}^{\ast }-u_{4}^{\ast }\right) ^{2}\right] }\right\rangle &\approx &1+%
\frac{1}{4}\left\langle \left( u_{1}-u_{3}\right) ^{2}+\left( u_{2}^{\ast
}-u_{4}^{\ast }\right) ^{2}\right\rangle =1, \\
\text{since \ }\left\langle u_{l}^{2}\right\rangle &=&\left\langle
u_{l}u_{k}^{\ast }\right\rangle =0\ ,\ \left( l\not=k\right)
\end{eqnarray*}%
whereas for small values of $\left\vert \mathbf{R}_{l+1}-\mathbf{R}%
_{l}\right\vert $ this exponential is always close to $1$ and the remaining
integration over angular variables can be perfomed in closed form (see
below). Thus, one can approximate the vertex part by the following simple
expression:%
\begin{equation}
\widetilde{\Gamma }_{4}\left( \mathbf{R}_{1},\mathbf{R}_{2},\mathbf{R}_{3},%
\mathbf{R}_{4}\right) =\frac{a_{x}\sqrt{\pi }}{2\pi }|c_{0}|^{4}e^{iq\left(
Z_{1}-Z_{2}+Z_{3}-Z_{4}\right) }e^{-\frac{1}{4}\sum |\rho _{l}|^{2}}
\label{Gamma4_fin}
\end{equation}%
which depends only on nearest neighboring coordinates.

Making use of Eq.(\ref{Gamma4_fin}), the remaining calculation of the
quartic term is similar to that used for the quadratic term, but
considerably massier. Below we present only an outline of the derivation.
Since integrations over $z_{i}$ are trivial we shall use from now on only 2D
vector notations with integrations over $Z_{i}$ written explicitly.

Combining Eqs.(\ref{Omega4_def}),(\ref{K4_def}),(\ref{Gamma4_fin}) our
starting expression for the quartic term is given by:%
\begin{equation*}
\Omega _{4}^{(s,d)}=\frac{k_{B}T}{V_{0}}\left( \frac{1}{2\pi \hbar \omega
_{c}}\right) ^{4}\frac{a_{x}\sqrt{\pi }}{2\pi L_{z}}|c_{0}|^{4}\sum_{\nu
}\int dZ_{1}dZ_{2}dZ_{3}dZ_{4}
\end{equation*}%
\begin{equation*}
\times \int \prod_{i=1}^{4}\frac{dk_{z,i}}{2\pi }%
e^{-ik_{z,1}(Z_{2}-Z_{1})}e^{ik_{z,2}(Z_{3}-Z_{2})}e^{-ik_{z,3}(Z_{4}-Z_{3})}e^{ik_{z,4}(Z_{1}-Z_{4})}e^{iq\left( Z_{1}-Z_{2}+Z_{3}-Z_{4}\right) }
\end{equation*}%
\begin{equation}
\ \ \times \int d^{2}\mathbf{r}_{1}d^{2}\mathbf{r}_{2}d^{2}\mathbf{r}%
_{3}d^{2}\mathbf{r}_{4}f\left( \mathbf{r}_{1}\right) f^{\ast }\left( \mathbf{%
r}_{2}\right) f\left( \mathbf{r}_{3}\right) f^{\ast }\left( \mathbf{r}%
_{4}\right) \times \Theta _{4}\left( \mathbf{r}_{1},\mathbf{r}_{2},\mathbf{r}%
_{3},\mathbf{r}_{4};\left\{ k_{z,i}\right\} ;\omega _{\nu }\right)
\label{Omega4_int}
\end{equation}%
where the function $\Theta _{4}\left( \mathbf{r}_{1},\mathbf{r}_{2},\mathbf{r%
}_{3},\mathbf{r}_{4};\left\{ k_{z,i}\right\} ;\omega _{\nu }\right) $
includes integration over all electron pair coordinates: \emph{\ }%
\begin{equation*}
\Theta _{4}\left( \mathbf{r}_{1},\mathbf{r}_{2},\mathbf{r}_{3},\mathbf{r}%
_{4};\left\{ k_{z,i}\right\} ;\omega _{\nu }\right) =\frac{1}{L_{x}L_{y}}%
\int d^{2}\mathbf{R}_{1}d^{2}\mathbf{R}_{2}d^{2}\mathbf{R}_{3}d^{2}\mathbf{R}%
_{4}e^{-\frac{1}{4}\sum |\rho _{l}|^{2}}\times
\end{equation*}%
\begin{equation*}
\int_{0}^{\infty }d\tau _{1}e^{-i\tau _{1}\left[ n_{F}-x_{1}^{2}-g-i%
\widetilde{\omega }_{\nu }-i\widetilde{\Gamma }\right] }\frac{\exp \left( 
\frac{R_{12}^{2}}{4}\frac{1+e^{-i\tau _{1}}}{1-e^{-i\tau _{1}}}\right) }{%
1-e^{i\tau _{1}}}\int_{0}^{\infty }d\tau _{2}e^{i\tau _{2}\left[
n_{F}-x_{2}^{2}+g+i\widetilde{\omega }_{\nu }+i\widetilde{\Gamma }\right] }%
\frac{\exp \left( \frac{R_{23}^{2}}{4}\frac{1+e^{i\tau _{2}}}{1-e^{i\tau
_{2}}}\right) }{1-e^{-i\tau _{2}}}\times
\end{equation*}%
\begin{equation}
\int_{0}^{\infty }d\tau _{3}e^{-i\tau _{3}\left[ n_{F}-x_{3}^{2}-g-i%
\widetilde{\omega }_{\nu }-i\widetilde{\Gamma }\right] }\frac{\exp \left( 
\frac{R_{34}^{2}}{4}\frac{1+e^{-i\tau _{3}}}{1-e^{-i\tau _{3}}}\right) }{%
1-e^{i\tau _{3}}}\int_{0}^{\infty }d\tau _{4}e^{i\tau _{4}\left[
n_{F}-x_{4}^{2}+g+i\widetilde{\omega }_{\nu }+i\widetilde{\Gamma }\right] }%
\frac{\exp \left( \frac{R_{41}^{2}}{4}\frac{1+e^{i\tau _{4}}}{1-e^{i\tau
_{4}}}\right) }{1-e^{-i\tau _{4}}}  \label{Theta4_def}
\end{equation}%
Here the coordinates, $\mathbf{R}_{i,i+1}$, in Eq.(\ref{Theta4_def}) are the
linear combinations of $\mathbf{\rho }_{l}=\mathbf{R}_{l+1}-\mathbf{R}_{l}$
and $\mathbf{\eta }_{l}=\mathbf{r}_{l+1}-\mathbf{r}_{l}$ :%
\begin{eqnarray}
\mathbf{R}_{12} &=&\mathbf{\rho }_{1}+\frac{1}{2}\mathbf{\eta }_{1};\ \ \ \
\ \ \ \ \ \ \mathbf{R}_{23}=\mathbf{\rho }_{2}-\frac{1}{2}\mathbf{\eta }_{2}
\notag \\
\mathbf{R}_{34} &=&\mathbf{\rho }_{3}+\frac{1}{2}\mathbf{\eta }_{3};\ \ \ \
\ \ \ \ \ \ \mathbf{R}_{41}=\mathbf{\rho }_{4}-\frac{1}{2}\mathbf{\eta }_{4}
\label{rel_coord}
\end{eqnarray}%
The gaussian integration over $\mathbf{\rho }_{l}$ reduces Eq. (\ref%
{Theta4_def}) to: 
\begin{equation*}
\Theta _{4}\left( \mathbf{r}_{1},\mathbf{r}_{2},\mathbf{r}_{3},\mathbf{r}%
_{4};\left\{ k_{z,i}\right\} ;\omega _{\nu }\right) =\int_{0}^{\infty }d\tau
_{1}d\tau _{2}d\tau _{3}d\tau _{4}\frac{\left( 2\pi \right) ^{3}}{\varkappa }%
\exp \left[ -\frac{1}{8\varkappa }\left( \mathbf{\eta }_{1}-\mathbf{\eta }%
_{2}+\mathbf{\eta }_{3}-\mathbf{\eta }_{4}\right) ^{2}\right]
\end{equation*}%
\begin{equation}
e^{-i\tau _{1}\left[ n_{F}-x_{1}^{2}-g-i\widetilde{\omega }_{\nu }-i%
\widetilde{\Gamma }\right] }e^{i\tau _{2}\left[ n_{F}-x_{2}^{2}+g+i%
\widetilde{\omega }_{\nu }+i\widetilde{\Gamma }\right] }e^{-i\tau _{3}\left[
n_{F}-x_{3}^{2}-g-i\widetilde{\omega }_{\nu }-i\widetilde{\Gamma }\right]
}e^{i\tau _{4}\left[ n_{F}-x_{4}^{2}+g+i\widetilde{\omega }_{\nu }+i%
\widetilde{\Gamma }\right] }  \label{Theta4}
\end{equation}%
where $\varkappa =4-e^{i\tau _{1}}-e^{-i\tau _{2}}-e^{i\tau _{3}}-e^{-i\tau
_{4}}$. It should be noted here that Eq.(\ref{Theta4}) has been obtained by
exploiting the fact that the dominant contributions to the integrals
originate in the regions where $\tau _{i}\ll 1$.

Furthermoe, noting that in the above equation the $\mathbf{\eta }_{i}$- and $%
k_{z}$-dependences are factorized, one can perfom the integrations over both
sets of variables separetely. For a $d$-wave superconductor we obtain: 
\begin{equation*}
\int d^{2}\mathbf{r}_{1}d^{2}\mathbf{r}_{2}d^{2}\mathbf{r}_{3}d^{2}\mathbf{r}%
_{4}f\left( \mathbf{r}_{1}\right) f^{\ast }\left( \mathbf{r}_{2}\right)
f\left( \mathbf{r}_{3}\right) f^{\ast }\left( \mathbf{r}_{4}\right) \times
\exp \left[ -\frac{1}{8\varkappa }\left( \mathbf{\eta }_{1}-\mathbf{\eta }%
_{2}+\mathbf{\eta }_{3}-\mathbf{\eta }_{4}\right) ^{2}\right]
\end{equation*}%
\begin{equation}
=\frac{1}{4^{3}}\left( 1-e^{-\frac{d^{2}}{\varkappa }}\right) ^{4}\left(
3+2e^{-\frac{d^{2}}{\varkappa }}+e^{-2\frac{d^{2}}{\varkappa }}\right)
^{2}\simeq \frac{9}{16}\frac{d^{8}}{\varkappa ^{4}}  \label{r_int}
\end{equation}%
where the last appoximation is valid under the same conditions discussed in
the derivation of the quadratic term. Thus the quartic term is transformed
to:%
\begin{equation*}
\Omega _{4}^{(d)}=\frac{k_{B}T}{\left( \hbar \omega _{c}\right) ^{4}}\frac{%
a_{x}\sqrt{\pi }}{\left( 2\pi \right) ^{3}}|c_{0}|^{4}\frac{9d^{8}}{%
16a_{H}^{8}}\sum_{\nu }\int dk_{z}\int_{0}^{\infty }\frac{d\tau _{1}d\tau
_{2}d\tau _{3}d\tau _{4}}{\left( 4-e^{i\tau _{1}}-e^{-i\tau _{2}}-e^{i\tau
_{3}}-e^{-i\tau _{4}}\right) ^{5}}
\end{equation*}%
\begin{equation}
e^{-i\left( \tau _{1}-\tau _{2}+\tau _{3}-\tau _{4}\right) \left( n_{F}-\xi
^{2}k_{z}^{2}-\xi ^{2}\left( \frac{q}{2}\right) ^{2}\right) }e^{-\left( \tau
_{1}+\tau _{2}+\tau _{3}+\tau _{4}\right) \left( \widetilde{\omega }_{\nu }+%
\widetilde{\Gamma }\right) }e^{i\left( \tau _{1}+\tau _{2}+\tau _{3}+\tau
_{4}\right) \left( g+\xi ^{2}k_{z}q\right) },  \label{Omega4_int2}
\end{equation}%
where an additional integration over $\zeta =\left( \tau _{1}-\tau _{2}+\tau
_{3}-\tau _{4}\right) /2$ for small $\tau _{i}$\ can be performed. Rescaling
variables as%
\begin{equation}
\varrho =\sqrt{n_{F}}\frac{\tau _{1}+\tau _{2}+\tau _{3}+\tau _{4}}{2};\ \ \
\ s=\frac{\sqrt{n_{F}}}{2}\left( \tau _{3}-\tau _{1}\right) ;\ \ \ \   \notag
\end{equation}%
and summing up over $\nu $ one obtain the final result for the quartic term:%
\begin{equation}
\Omega _{4}^{(d)}=B^{(d)}\int_{0}^{\infty }d\varrho \frac{1-e^{-\frac{%
2\omega _{D}}{\sqrt{\mu \hbar \omega _{c}}}\varrho }}{\sinh \left( \frac{%
2\pi k_{B}T}{\sqrt{\mu \hbar \omega _{c}}}\varrho \right) }e^{-\frac{2%
\widetilde{\Gamma }}{\sqrt{n_{F}}}\varrho }\cos \left( 2\varrho g_{0}\right)
\Theta _{4}^{\left( d\right) }\left( \varrho ,q\right)  \label{Omega4_fin}
\end{equation}%
where $B^{(d)}=c_{4}^{(d)}B_{0}$ with $c_{4}^{(d)}=\frac{3}{16}\left( \frac{%
k_{F}d}{\pi }\right) ^{8}$, $B_{0}=\left( \frac{\sqrt{\pi }}{a_{x}}\right) 
\frac{\pi k_{B}T\Delta _{0}^{4}}{\left( \mu \hbar \omega _{c}\right) ^{3/2}}%
N\left( 0\right) $, and 
\begin{equation}
\Theta _{4}^{\left( d\right) }\left( \varrho ,q\right) =\int_{0}^{1}dv\left(
1-v^{2}\right) ^{4}e^{-\frac{1}{2}\varrho ^{2}\left( 1-v^{2}\right) }\cos
\left( 2vq\varrho \right) \left( \int_{0}^{\varrho }dse^{-s^{2}\left(
1-v^{2}\right) }\right) ^{2}  \label{Theta4_fin}
\end{equation}%
\end{widetext}
The result for an $s$-wave superconductor can be obtained from Eq.(\ref%
{Omega4_fin}) by replacing the factor $\left( 1-v^{2}\right) ^{4}$ in the
definition of $\Theta _{4}^{\left( d\right) }$ and the factor $c_{4}^{(d)}$
in the normalization coefficient $B^{(d)}$ with unity. The $s$-wave quartic
term for zero spin splitting is equivalent to that obtained in Ref.\cite%
{MRVW92}.

\section{Results and Discussion}

The analysis presented in the previous sections enables us to write a
GL-like expansion of the SC contribution to the thermodynamic potential for
an $s$ or $d_{x^{2}-y^{2}}$- wave pairing up to second order in $\Delta
_{0}^{2}$:%
\begin{eqnarray}
\Omega ^{\left( s,d\right) } &=&\alpha ^{\left( s,d\right) }\left(
t,b,q\right) \Delta _{0}^{2}+\frac{1}{2}\beta ^{\left( s,d\right) }\left(
t,b,q\right) \Delta _{0}^{4}  \notag \\
&+&\frac{1}{3}\gamma ^{\left( s,d\right) }\left( t,b,q\right) \Delta
_{0}^{6}+...  \label{omega}
\end{eqnarray}

For the quadratic term we have: 
\begin{eqnarray}
\alpha ^{\left( s,d\right) }\left( t,b,q\right) &=&\frac{1}{\lambda }-\frac{%
c_{2}^{\left( s,d\right) }}{\varsigma \left( T\right) }\int_{0}^{\infty
}d\rho \frac{\left( 1-e^{-\frac{2\rho }{\varsigma \left( T_{D}\right) }%
}\right) }{\sinh \left( \frac{2\rho }{\varsigma \left( T\right) }\right) } 
\notag \\
&&e^{-2\rho /l}\cos \left( \frac{2g}{r_{F}}\rho \right) \Theta _{2}^{\left(
s,d\right) }\left( \rho ,q\right)  \notag \\
\Theta _{2}^{\left( s,d\right) }\left( \rho ,q\right)
&=&\int_{0}^{v_{0}}d\nu \vartheta _{2}^{\left( s,d\right) }\left( v\right)
\cos \left( q\rho v/\chi _{a}\right)  \notag \\
&&\exp \left[ -\left( 1-v^{2}\right) \rho ^{2}/2a_{H}^{2}\right]
\label{alpha}
\end{eqnarray}%
where $\lambda =N(0)V$ , $\ \varsigma \left( T_{D}\right) \equiv \hbar
v_{F}/\pi k_{B}T_{D}$ , with $T_{D}$ the Debye temperature, and $v_{F}=\sqrt{%
2\varepsilon _{F}/m^{\ast }}$-the inplane Fermi velocity, $\varsigma \left(
T\right) \equiv \hbar v_{F}/\pi k_{B}T$- the thermal mean-free path, $r_{F}=%
\sqrt{2n_{F}}a_{H}$- the electronic cyclotron radius at the Fermi energy,
and $l$ is the mean-free path due to impurity scattering.

The differences between $s$-wave and $d$-wave SCs are given by $%
c_{2}^{\left( s\right) }=1$, $\ c_{2}^{\left( d\right) }=3\left( \frac{k_{F}d%
}{\pi }\right) ^{4}$, and $\vartheta _{2}^{\left( s\right) }\left( v\right)
=1$, $\vartheta _{2}^{\left( d\right) }\left( v\right) =\left(
1-v^{2}\right) ^{2}$.

The quartic term has a similar structure:%
\begin{eqnarray}
&&\beta ^{\left( s,d\right) }\left( t,b,q\right) =B_{0}c_{4}^{\left(
s,d\right) }\int_{0}^{\infty }d\rho \frac{\left( 1-e^{-\frac{2\rho }{%
\varsigma \left( T_{D}\right) }}\right) }{\sinh \left( \frac{2\rho }{%
\varsigma \left( T\right) }\right) }  \notag \\
&&e^{-2\rho /l}\cos \left( \frac{2g}{r_{F}}\rho \right) \Theta _{4}^{\left(
s,d\right) }\left( \rho ,q\right)  \label{beta} \\
&&\Theta _{4}^{\left( s,d\right) }\left( \rho ,q\right)
=\int_{0}^{v_{0}}d\nu \vartheta _{2}^{\left( s,d\right) }\left( v\right)
\cos \left( q\rho v/\chi _{a}\right)  \notag \\
&&\exp \left[ -\left( 1-v^{2}\right) \rho ^{2}/4a_{H}^{2}\right] \left(
\int_{0}^{\rho /\sqrt{2}a_{H}}dse^{-s^{2}\left( 1-v^{2}\right) }\right) ^{2}
\notag
\end{eqnarray}%
where \ $B_{0}=\left( \frac{\sqrt{\pi }}{a_{x}}\right) \frac{N\left(
0\right)\pi k_{B}T}{\left( \varepsilon _{F}\hbar \omega _{c}\right) ^{3/2}}$
and $c_{4}^{\left( s\right) }=1$, $c_{4}^{(d)}=\frac{3 }{16} \left( \frac{%
k_{F}d}{\pi }\right) ^{8}$, $\vartheta _{4}^{\left( s\right) }\left(
v\right) =1$, and $\vartheta _{4}^{\left( d\right) }\left( v\right) =\left(
1-v^{2}\right) ^{4}$. \ 

On the basis of the above formulas we discuss below the H-T phase diagram
for different values of the relevant parameters. Three independent
dimensionless parameters:$\ \left( 2a_{H}/r_{F}\right) g$ , $%
2a_{H}/\varsigma \left( T\right) $ and $qa_{H}/\chi _{a}$, control the basic
integrals in these equations. The first two parameters measure the strength
of the spin and thermal pair-breaking mechanisms, respectively, relative to
the orbital (diamagnetic) depairing. The third parameter determines the
relative strength of the compensating FFLO mechanism. The value of the spin
pair-breaking parameter, $\ \sigma \equiv g\left( 2a_{H}/r_{F}\right)
_{H=H_{c20}^{orb}}$ , where $H_{c20}^{orb}$ is the upper critical field at $%
T=0$ in the absence of spin pair-breaking, is related to the well known Maki
parameter \cite{Maki64}, $\alpha _{M}=\frac{\left( \hbar e/m_{0}c\right)
H_{c20}^{orb}}{1.76k_{B}T_{c0}}$, by: $\sigma =1.1\alpha _{M}$. Here $T_{c0}$
is the transition temperature at zero magnetic field.

%

\begin{figure}[tbp]
\includegraphics[width=7cm]{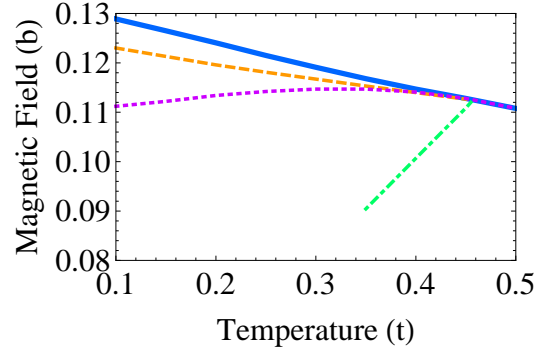}
\caption{Phase transition lines for a 3D system with s-wave pairing : N-SC
transitions for order parameters with (solid line) or without (doted line)
FFLO modulation, and FFLO-BCS transitions obtained for GL free energy with
(dashed line) or without (doted-dashed line) quartic correction. The N-SC
transition is of second order whereas the FFLO-BCS transition is of first
order. The value of the spin splitting parameter is $\protect\sigma=1.8$.}
\label{fig:1}
\end{figure}

\begin{figure}[tbp]
\includegraphics[width=7cm]{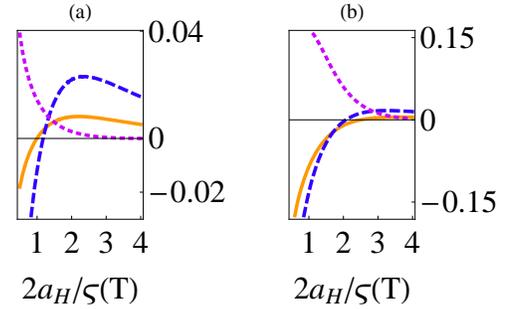}
\caption{ The GL coefficients (in arbitrary units ) $\protect\beta \left(
t,b,q=0\right) $ (dashed lines), $\frac{d\protect\alpha \left(
t,b,q=0\right) }{dq^{2}}$(solid lines) ,$\frac{d\protect\beta \left(
t,b,q=0\right) }{dq^{2}}$(doted lines) as functions of the parameter $\frac{%
2a_{H}}{\protect\varsigma \left( T\right) }$ for (a) $v_{0}=.4$ \ and (b) $%
v_{0}=1$. }
\label{fig:2}
\end{figure}

As we shall show below, the situation $T_{fflo}>T^{\ast }$, where $T^{\ast }$
is the temperature at which $\beta \left( t,b_{c2},q=0\right) =0$ , is
realized in 3D systems (corresponding to $v_{0}=1$), regardless of the
spin-splitting strength and the type of electron pairing. \ A typical phase
diagram is shown in Fig. 1 for $s$-wave pairing and spin pair-breaking
parameter $\sigma =3$.

\begin{figure}[tbp]
\includegraphics[width=6cm]{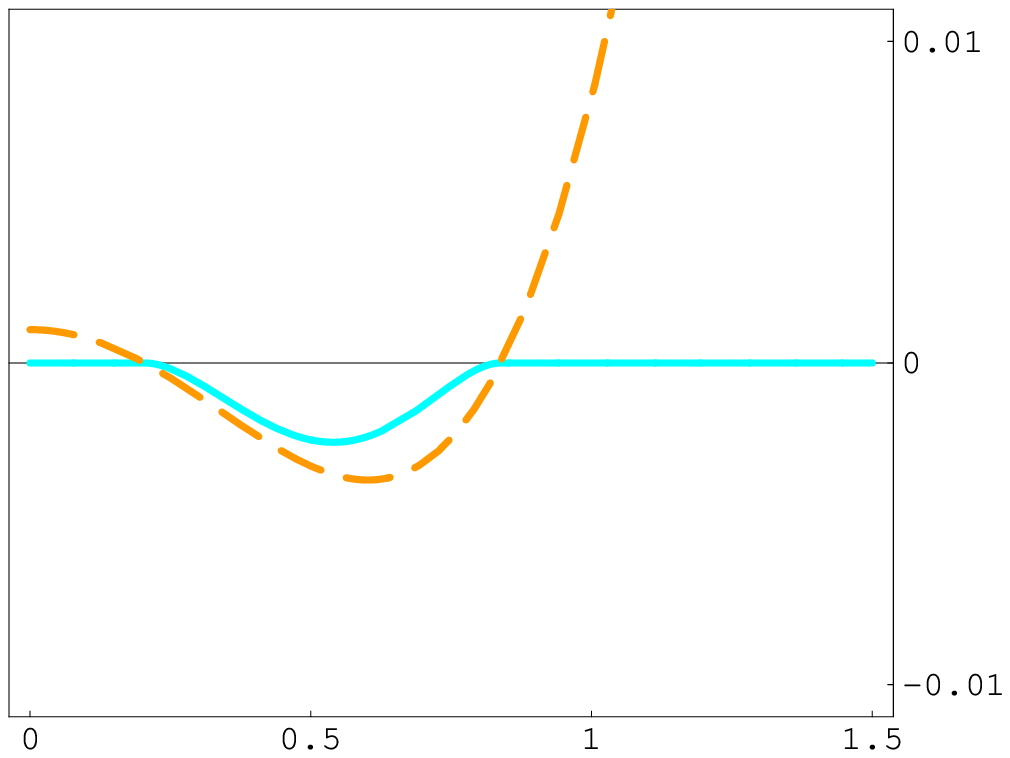} \\
\includegraphics[width=6cm]{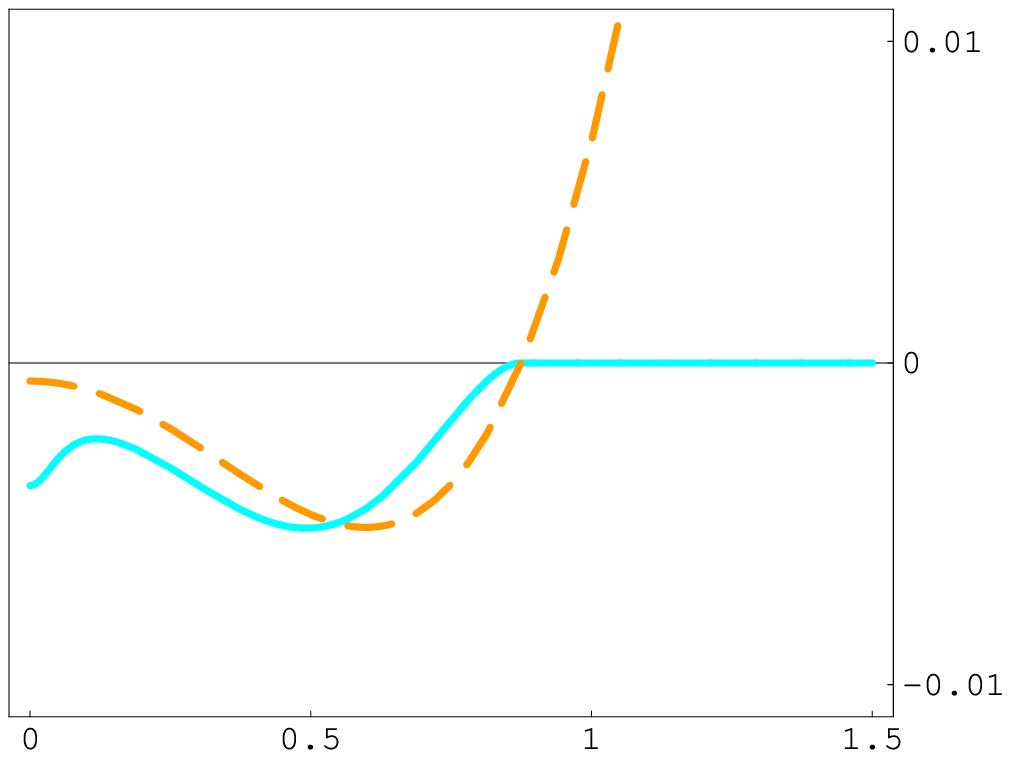} \\
\includegraphics[width=6cm]{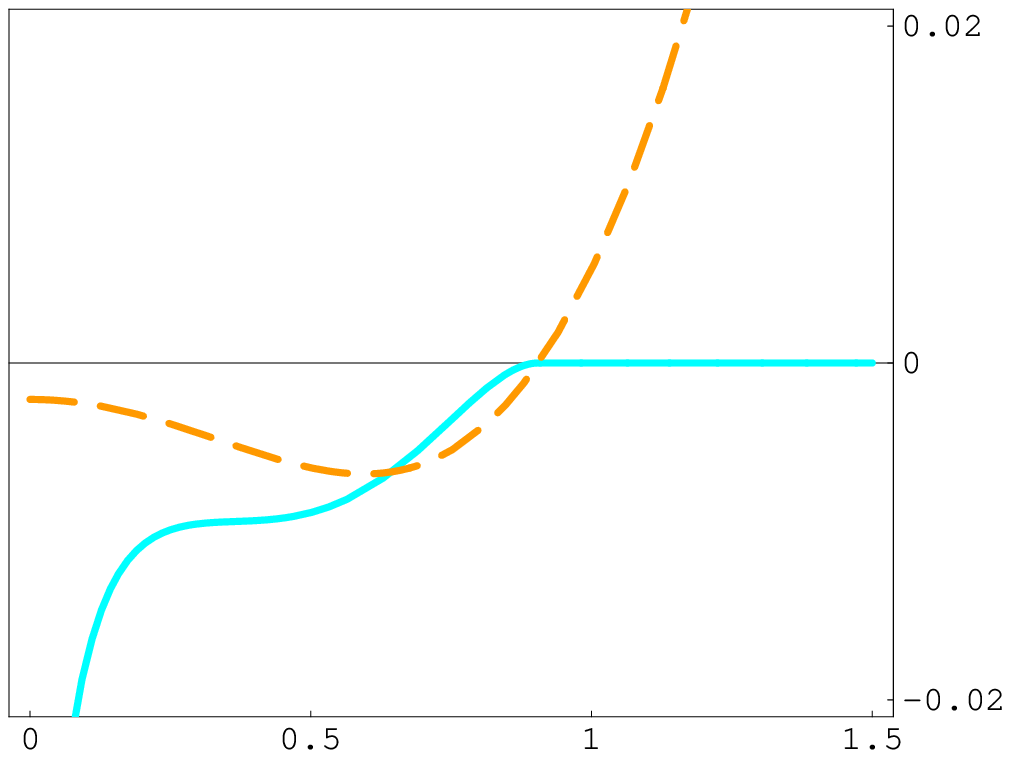} \\
\hspace{0cm}  $\widetilde{q}$
\caption{(a) The dependence of the GL coefficient $\protect\alpha $ (orange)
and the mean field SC free energy, $-\protect\theta \left( \protect\alpha %
\right) \frac{\protect\alpha ^{2}}{2\protect\beta _{A}\protect\beta }$
(blue), on the modulation paramer, \ $\widetilde{q}\equiv \left( \frac{2}{%
\protect\sigma }\right) \protect\sqrt{\frac{m^{\ast }}{m_{z}^{\ast }}}\left(
qa_{H_{c20}^{orb}}\right) $, in a 3D system ($v_{0}=1$) , at $t=0.4$, $%
b=0.1142$, i.e. near the tricritical point, just below the Normal-nonuniform
SC transition. It is seen that $\protect\alpha <0$ in a small region around $%
\ \widetilde{q}=0.6$ where the SC free energy has a minimum.{\protect\large %
\ }The value of the spin splitting parameter is $\protect\sigma =3$. (b):
The same as in (a) but for a slightly lower field, $b=0.114$, where a
uniform ($q=0$) metastable SC state is present. (c): The same as in (b) but
for a slightly lower field, $b=0.1139$, where a uniform ($q=0$)\ equilibrium
SC state is present, while a metastable SC state exists at $q\not=0$.}
\label{fig:3}
\end{figure}

As long as $T>T^{\ast }$ (so that $\beta \left( t,b_{c2},q=0\right) >0$) the
normal to SC (N-SC) phase transition is of second-order and the (reduced)
critical field, $b_{c2}\left( t\right) $, can be determined as the maximal
value of $b\equiv H/H_{c20}^{orb}$ obtained from the equation $\alpha \left(
t,b,q\right) =0$ for all values of $q$ , at the (reduced) temperature $%
t\equiv T/T_{c0}$. The solution of this equation for $q=0$ yields a
transition line, $b=b_{c2}^{\left( 0\right) }\left( t\right) $, ignoring the
possibility of a FFLO state. The tricritical point, $T_{fflo}$, is defined
as the maximal temperature at which $b_{c2}\left( t\right) >b_{c2}^{\left(
0\right) }\left( t\right) $. It can alternatively be determined from the
equation $\frac{d\alpha \left( t,b_{c2},q=0\right) }{dq^{2}}=0$ , which is
equivalent to the condition for vanishing of the coefficient of $\left\vert
\nabla \Delta \right\vert ^{2}$ in a gradient expansion of the SC\ free
energy\cite{houzet01}.

For $T<T^{\ast }$ and sufficiently strong spin pair-breaking there can be a
changeover to first-order SC transitions, but since $\beta \left(
t,b_{c2},q\neq 0\right) >\beta \left( t,b_{c2},q=0\right) $ (see Fig.2 ),
the segment of the $b_{c2}\left( t\right) $-line with first order
transitions arises only at very low temperatures. For moderate\ $\sigma $
values\ the coefficient\ $\beta \left( t,b_{c2},q\right) $\ at optimal\ $q$\
is always positive and the N-SC transition is of the second order at
arbitrarily low temperature.

The transition within the SC region from the nonuniform (FFLO) to uniform
(BCS) phase at $T>T^{\ast }$ can not be obtained just by analyzing the
quadratic term $\alpha \left( t,b,q\right) $ since the SC order parameter is
finite there. It can be obtained by minimizing the SC free energy (including
both quadratic and quartic terms) with respect to the modulation wave number 
$q$. \ Neglecting the sixth and higher order terms in the expansion, the
corresponding (standard) GL free energy, $\Omega \left( q\right) \simeq
-\theta \left( \alpha \right) \frac{\alpha ^{2}}{2\beta }$, ( $\theta \left(
\alpha \right) $\ being the Heaviside step function), which has a single
minimum at $q\not=0$ for field near $b_{c2}$ (see Fig. 3a), developes a
double-well structure (see Fig. 3b) as a function of $q$ upon decreasing the
field below $b_{c2}$ at a given temperature $T$ (due to the symmetry $%
q\leftrightarrow -q$ only positive values may be considered). One of these
minima is always at $q=0$, and it becomes energetically favorable at a
critical field for a first order phase transition from the FFLO to the
uniform BCS phase. The second (metastable) minimum at $q\not=0$\ disappears
completely upon further field decrease (see Fig. 3c).\ \ \

\begin{figure}[th]
\includegraphics[width=6cm]{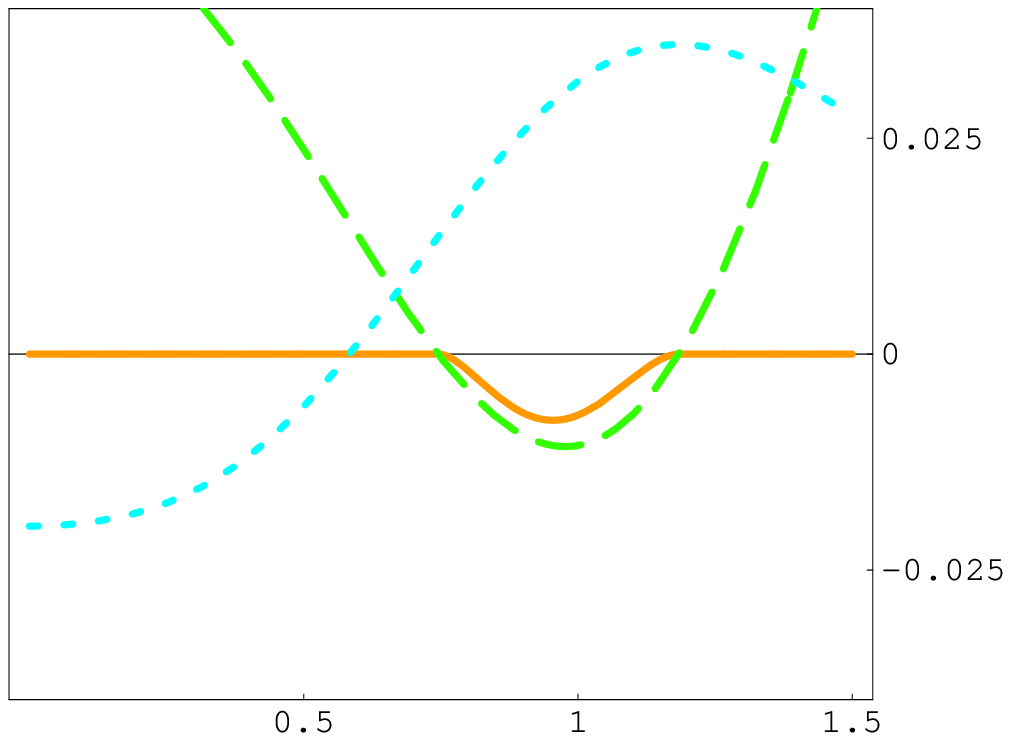} \\
\includegraphics[width=6cm]{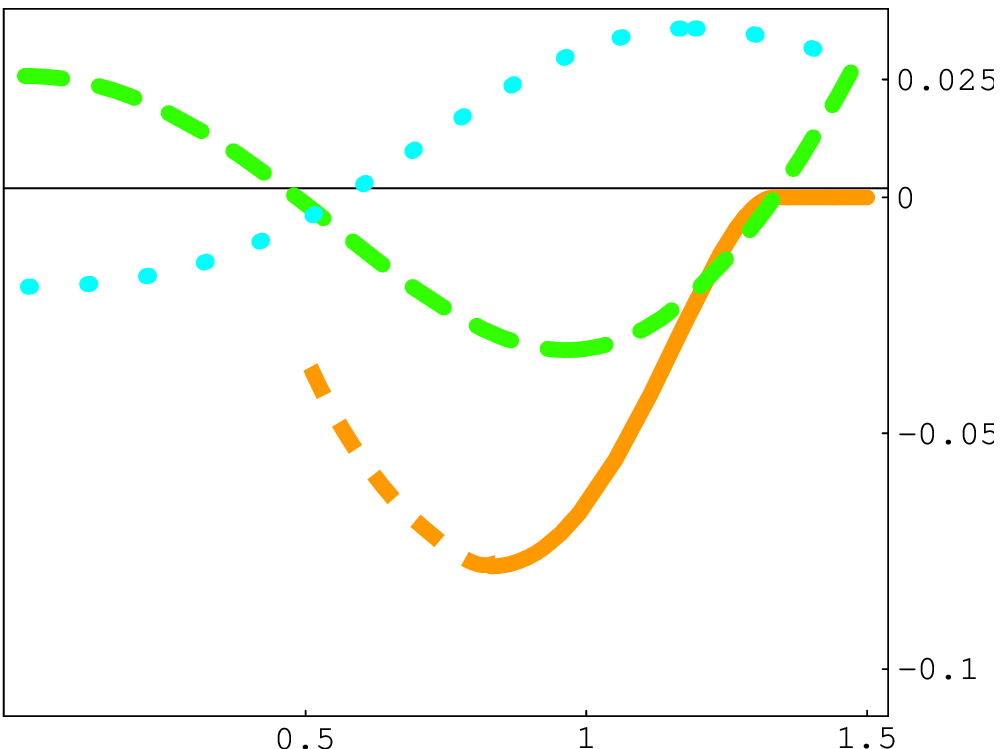}  \\
$\widetilde{q}$
\caption{The GL coefficients, $\protect\alpha $\ (green) and $\protect\beta $%
\ (blue), and the mean field SC free energy (orange) as functions of the
modulation parameter, $\widetilde{q}$, in a 3D system ($v_{0}=1$) at a
relatively low temperature $t=.25$\ and decreasing field values (a) $b=.118$%
\ and (b) $b=.117$. Note the vanishing of $\protect\beta $ inside the region
where $\protect\alpha <0$ , around which the used approximation, $-\protect%
\theta \left( \protect\alpha \right) \frac{\protect\alpha ^{2}}{2\protect%
\beta _{A}\protect\beta }$, for $\Omega (q)$ breaks down (dashed sector of
the orange line ). }
\label{fig:4}
\end{figure}

At temperatures $T$ below $T^{\ast }$ the first two terms in the expansion
of the thermodynamic potential are not sufficient to correctly describe the
uniform SC state since for negative $\beta $-\ values the scale of the SC
free energy is determined by the sixth order term.\texttt{\ }In contrast,
the free energy of the nonuniform state, where $\beta \left(
t,b,q\not=0\right) >0$, can be obtained from the stantard GL functional
(with the assumption that the contribution of the sixth order term is small
compared to that of the quartic term). The characteristic $q$-dependences of
the GL coefficients, $\alpha $\ and $\beta $, and the mean field free energy 
$-\theta \left( \alpha \right) \frac{\alpha ^{2}}{2\beta _{A}\beta }$, for $%
T<T^{\ast }$\ are illustrated in Fig. 4.\texttt{\ }Whereas at high fields
(Fig. 4a) the minimum of the SC energy occurs in a region where $\beta >0$,
at lower fields (see Fig. 4b) it approaches the expanding temperature domain
of negative $\beta $.\texttt{\ }Thus, even for moderate spin splitting and
low temperature the transition line from the nonuniform to uniform SC state
cannot be determined without knowing the sixth-order term. It is clear,
however, that this transition is of the first order.

It should be noted that if one attempts to determine the FFLO-BCS phase
boundary from the equation $\frac{d\alpha \left( t,b,q=0\right) }{dq^{2}}=0$
it will greatly overestimate the size of the FFLO phase as compared to that
obtained by minimizing $-\frac{\alpha ^{2}}{2\beta }$ (see Fig. 1). This
remarkable difference is due to the strong $q^{2}$-dependence of the quartic
coefficient $\beta $ (see Fig.2).

The suppression of the orbital effect in the considered 3D systems, with
ellipsoidal Fermi surfaces contained entirely within the BZ, is due to the
factor $1-v^{2}$ appearing in the Gaussian exponents of Eqs. (\ref{alpha}),(%
\ref{beta}). The recovery of this effect in quasi-2D systems with truncated
ellipsoidal Fermi surface, where $v_{0}<1$, can reverse the relation between 
$T_{fflo}$ and $T^{\ast }$. Fig. 2b, where the GL coefficients are shown for 
$\sigma =1.8$, $v_{0}=.4$ and $s$-wave pairing, illustrates the situation
with $T_{fflo}<T^{\ast }$, which occurs for all values of $v_{0}$ below a
critical dimensionality $v_{0,cr}\approx 0.44$ (see Fig. 5), and depends
only weakly on the spin-splitting parameter $\sigma $.

\begin{figure}[th]
\includegraphics[width=6cm]{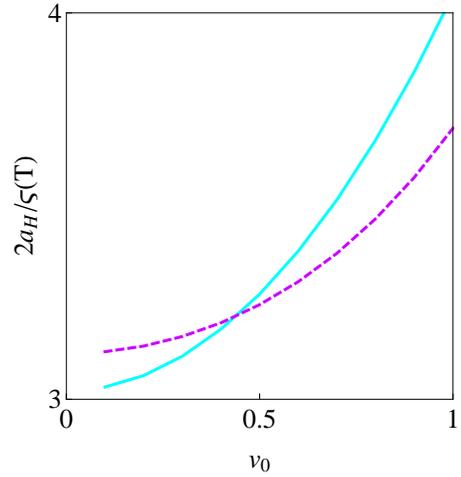}
\caption{Solutions$\ \frac{2a_{H}}{\protect\varsigma \left( T\right) }%
\propto T$ of the equation $\protect\beta \left( q=0\right) =0$,
corresponding to $T^{\ast }$ (dashed line), and the equation $\frac{d\protect%
\alpha \left( q=0\right) }{dq^{2}}=0$, corresponding to $T_{fflo}$ (solid
line) vs the dimensionality crossover parameter $v_{0}$ for $\protect\sigma %
=2.5$.}
\label{fig:5}
\end{figure}

\begin{figure}[tbp]
\includegraphics[width=7cm]{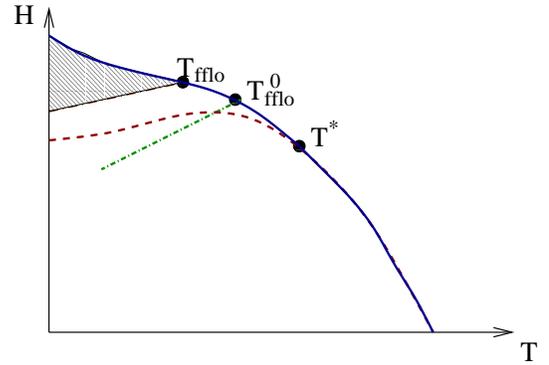}
\caption{Schematic phase diagram for a quasi 2D system. The shaded area
corresponds to a nonuniform SC (FFLO) phase, the dashed line corresponds to $%
\protect\alpha \left( t,b,q=0\right) =0$ , and the dotted-dashed line can be
obtained from $\frac{d}{dq^{2}}\protect\alpha \left( t,b,q=0\right) =0$.}
\label{fig:6}
\end{figure}

The corresponding phase diagram (see Fig. 6 ) for $v_{0}$ below this
crossing point is quite different from that found for the 3D systems shown
in Fig.1. First of all, since $\beta <0$ , one may use Eqs. (\ref{alpha}),(%
\ref{beta})\ to determine the phase diagram only under the assumption that
the sixth order coefficient $\gamma $ is positive (see Ref.\cite{adachi03}).
In this case a discontinuous SC transition occurs at $\Omega \left( \Delta
_{0}^{2}\right) =0$\ with $\Delta _{0}^{2}=\left( \frac{3\left\vert \beta
\right\vert }{4\gamma }\right) $ and $\alpha =\frac{3\beta ^{2}}{16\gamma }%
>0 $ , and the corresponding critical field, $b_{c2}(t)$ , should be larger
than $b_{c2}^{0}\left( t\right) $ , obtained from the equation $\alpha
(q=0)=0$. \ Thus, at a temperature below $T^{\ast }$, the N-SC phase
boundary includes a segment of first order transitions, which may end at
zero temperature, or at a finite temperature, depending on the
spin-splitting strength. This dependence appears because of the competition
between the decreasing explicit dependence of $\beta $ on decreasing
temperature and its increasing implicit dependence through $q\left( T\right) 
$ at the FFLO state. \ The boundary between the BCS and FFLO states should
be determined by minimizing the free energy, Eq. (\ref{omega}), with respect
to $q$. This may be restricted to the explicit dependence on $q$ since the
order parameter is determined by: $\frac{\partial \Omega }{\partial \Delta
^{2}}=0$. Consequently the positive sign of $\frac{d\beta \left(
t,b,q=0\right) }{dq^{2}}$ (see Fig. 2b) results in partial cancellation of
the leading contribution to $\frac{\partial \Omega }{\partial q^{2}}$, which
is proportional to $\frac{d\alpha \left( t,b,q=0\right) }{dq^{2}}$\ and
negative in the FFLO part of the phase diagram.\ Moreover, since for the
discontinuous transition the order parameter is finite just below the
transition the higher order terms in $\Delta _{0}^{2}$ (Eq. (\ref{omega}))
should be taken into account. As a result $T_{fflo}$ should be smaller than $%
T_{fflo}^{0}$- the temperature obtained from the equation $\frac{d\alpha
\left( t,b,q=0\right) }{dq^{2}}$ $=0$ , as schematically shown in Fig. 6.

\section{ Conclusions}

It is shown that the expected changeover to first-order SC transitions in
clean, strongly type-II superconductors in the Pauli paramagnetic limit can
take place only in materials with quasi-cylindrical Fermi surfaces,
regardless of the type of the electron (s or d-wave) pairing interaction
which leads to superconductivity. This finding clarifies the confusing
current literature on this topic\cite{gruenberg66},\cite{houzet01},\cite%
{adachi03}.

The observation of such a changeover in the heavy fermion compound CeCoIn$%
_{5}$ for magnetic field orientation perpendicular to the easy conducting
plane\cite{Bianchi0203} is consistent with the quasi-2D character of its
electronic band structure \cite{Settai01}. The interesting situation of a 2D
superconductor under a magnetic field parallel to the conducting plane, for
which a changeover to discontinuous SC transitions was reported very recently%
\cite{Lortz07}, is more subtle since the vanishingly small cyclotron
frequency characterizing this case does not allow utilization of the Landau
orbitals approach employed here (for a recent review see, e.g.\cite%
{Matsuda-Shimahara07}).

\textbf{Acknowledgements}: This research was supported by the Israel Science
Foundation founded by the Academy of Sciences and Humanities, by the
Argentinian Research Fund at the Technion, and by EuroMagNET under the EU
contract RII3-CT-2004-506239.

\end{document}